\documentstyle[prl,twocolumn,aps]{revtex}
\input{epsf}

\newcommand{\bq}{\begin{equation}}
\newcommand{\ee}{\end{equation}}
\newcommand{\fr}[2]{\frac{#1}{#2}}
\newcommand{\eps}{\varepsilon}

\begin{document}
\draft

\title{
Towards an Explanation of the Mesoscopic Double-Slit Experiment,\\
a new model for charging of a Quantum Dot }

\author{P.G.Silvestrov$^{1,2}$ and Y.Imry$^{2}$}

\address{$^{1}$Budker Institute of Nuclear Physics, 630090
Novosibirsk, Russia,\\
$^{2}$ Weizmann Institute of Science, Rehovot 76100, Israel}

\maketitle
\date{today}

\begin{abstract}

For a quantum dot (QD) in the intermediate regime between
integrable and fully chaotic, the widths of single-particle
levels naturally differ by orders of magnitude.  In particular,
the width of one strongly coupled level may be larger than the
spacing between other, very narrow, levels. In this case many
consecutive Coulomb blockade peaks are due to occupation of the
same broad level. Between the peaks the electron jumps from this
level to one of the narrow levels and the transmission through
the dot at the next resonance essentially repeats that at the
previous one. This offers a natural explanation to the recently
observed behavior of the transmission phase in an interferometer
with a QD.

\end{abstract}


In spite of much progress in the fabrication and experimental
investigation of ultrasmall few-electron devices - such as
Quantum dots \cite{LKCM}, many experimentally observed features
of these systems still remain unexplained.

A challenging problem which has resisted adequate theoretical
interpretation arises from the experiment \cite{Heiblum2} which
determines the phase of the wave transmitted through the QD
\cite{endnote}. The main goal of this paper will be to find a
mechanism which may lead to a satisfactory explanation of these
results. Hopefully our approach will also allow to shed some
light on other open problems concerning the Coulomb
blockade~(CB)~\cite{CB} in QD-s.

In the experiment of ref.~\cite{Heiblum2}, in addition to the
conductance of the QD, the phase of the electron transmitted
through the QD was measured via an interference arrangement.  In
accordance with the Breit-Wigner picture, the phase increased by
$\pi$ around each CB peak. Absolutely unexpected, however, was a
fast jump of the phase by $-\pi$ between the resonances near the
minimum of the transmitted current.  Such a behavior is in
evident contradiction with what one would expect if the
transmission of the current proceeds via consecutive levels in
1-dimensional quantum well.

In a two--dimensional QD the phase drops associated with the
nodes of the transmission amplitude arise already within the
single--particle picture~\cite{Deo,Fano}. However, in order to
have a sequence of such events one should consider a QD of a
very special form.  The model of ref.~\cite{Oreg} also does not
allow to explain the series of drops.  The mechanism of
refs.~\cite{Hackenbroich,Gefen} makes nontrivial assumptions on
the geometry of the QD and the way it changes under the change
of plunger gate voltage. An interesting generic mechanism
suggested recently in ref.~\cite{Baltin} may indeed lead to the
correlations in transmission at many consecutive valleys, but
the predicted phase behavior differs from what has been seen
experimentally.

In this paper we propose a mechanism according to which the
transmission at many CB peaks proceeds through one and the same
level in the QD. This means that the phases at the wings of
different resonances should coincide and the increase by $\pi$
at the resonance must be compensated. This compensation occurs
via narrow jumps between the resonances and is accompanied by a
fast rearrangement of the electrons in the dot.

Although the experiment~\cite{Heiblum2} was clearly done in the
CB regime, the widths of the resonances turned out to have been
anomalously large, only few times smaller than the charging
energy. Also the widths and heights of all observed resonances
are very similar.  These surprising features of the results of
ref.~\cite{Heiblum2}, which have not attracted so wide an
attention as the phase jumps, also find natural explanation
within our picture.  Our mechanism requires the QD not to be
fully chaotic (neither do we require an integrable QD).  It is
not clear, how chaotic was the dot used in the experiment.
However, the QD containing $\sim 200$ electrons was $\sim 50$
times smaller than the nominal elastic mean free path. Thus,
disorder should not have been essential for the dynamics of the
electrons.

It is generally believed that the CB is observed only if the
widths of resonances are small compared to the single-particle
level spacing in the dot $\Delta$.  This condition assumes that
couplings of all levels to the leads are of the same order of
magnitude. However, as we will show at the end of this paper,
even for nonintegrable ballistic QD-s the widths of the
resonances may vary by orders of magnitude. In this case it does
not make sense to compare the width of few broad resonances with
the level spacing, determined by the majority of narrow,
practically decoupled, levels.

A useful theoretical model for the description of charging
effects in QD-s is the tunneling Hamiltonian in the constant 
interaction ($U_{CB}$) approximation (see e.g.~\cite{CB}) 
\begin{eqnarray}\label{Ham}
&\,&H=\sum_i \eps_i c^+_i c_i + U_{CB}\sum_{i<j} c^+_i c_i c^+_j
c_j  
\\ &\,& \ \ \ \   +   \sum_k\eps(k) a^{+}_k a_k +
\sum_{k,i}[t_i c^+_i a_k +h.c.]
\ . \nonumber
\end{eqnarray}
Here $c(c^+)$ and $a(a^+)$ are the annihilation(creation)
operators for electrons in the dot and in the lead and $\eps_i,
\eps(k)$ are the single-particle energies. We do not introduce
the $k$ dependence of the tunneling matrix elements $t_i$.
Since our approach is mainly based on the energetics of the QD,
it is enough to consider only one lead.  Summation over spin
orientations is easily included.  Also under the assumption of
capacitive coupling to the gate, the levels in the dot flow
uniformly with the voltage
\bq\label{1}
\eps_i=\eps_i{(V_g=0)} -V_g  \ .
\ee
The energies of the electrons in the wire are given by
\bq\label{11}
\eps(k)={k^2}/{2m}-E_F \ .
\ee
Here $k={n\pi}/{L}$, a (very)large $n$ is the level number 
in the wire and $L$ is the length of the wire.

For our purposes, it will be possible to simplify further the
Hamiltonian~(\ref{Ham}). We will consider the case where the
coupling of one particular level $N$ is dominant $t_N\gg t_i$,
$i\ne N$. If the width of this level is larger than the
single-particle level spacing $\Delta$, a very nontrivial regime
of charging of the QD may be described by means of second-order
perturbation theory estimates.  Surprisingly, this simple limit
of CB has not been considered yet.

An example of a system for which the widths differ drastically
is the integrable QD\cite{Hackenbroich,Gefen}.  However, it is
hard to believe that the large ($N_e\sim 100 \div 1000$) QD may
be even close to integrable.  Nevertheless, at least in
classical mechanics, a considerable gap is left between
integrable and fully chaotic systems. Even in a nonintegrable
dot two kinds of trajectories - quasi-periodic and chaotic  may
coexist. In this case, in 2-dimensions any trajectory (even a
chaotic one!) does not cover all the phase space  allowed by
energy conservation.  Consequently, the corresponding wave
functions do not cover all the area of the QD. If such a regime
is realized in QD-s, it easily explains why the widths of the
resonances may vary by orders of magnitude. Moreover, many other
features of such a QD may differ strongly from those of the
chaotic QD~\cite{Stopa}.  An explicit numerical example, which
supports the existence of such regime will be given later.

Now we turn to the many--particle effects arising for the 
Hamiltonian~(\ref{Ham}) in the case of only one
($N$-th) level in the dot coupled strongly to the wire
\bq\label{2}
\Gamma\equiv \Gamma_N = 2\pi |t_N|^2 {dn}/{d\eps} \gg \Delta \ . 
\ee
(Here $n$ is the same as in the eq.~(\ref{11}) and
${dn}/{d\eps}$ is taken at the Fermi energy $\eps=0$.) The
widths of the other levels are taken to be much smaller than the
level spacing and may be neglected in the first approximation.
The charging energy is still  very large $U_{CB}\gg\Gamma$. We
shall show that transmission of a current at about
$(\Gamma/\Delta)\ln(U_{CB}/\Gamma)$ consecutive CB peaks will
proceed through one and the same level $\eps_N$.

Let the levels with $i\le 0$ in the QD be occupied. Our aim is
to find the total energy $E_{tot}$ of the true ground state of
the dot at different values of $V_g$.  Without loss of
generality we may assume that the summation over $i$ in
eq.~(\ref{Ham}) goes only over $i>0$. (Thus we subtract from the
total energy the trivial constant corresponding to
selfinteraction of electrons with $i\le 0$. Coulomb interaction
between electrons at the levels with $i\le 0$ and $i>0$ is
included into $\eps_{i>0}$.) Also let us subtract from the total
energy the trivial energy of electron gas in the leads
$\sum\eps(k)$.

Let us consider spinless electrons. For large positive
$\eps_N(V_g)\gg \Gamma$ the only contribution to the total
energy $E_{tot}$ is given by the second order correction (the
levels in the wire are lowered due to the repulsion from the
unoccupied level $\eps_N$)
\bq\label{3}
E^{(0)}_{tot}=\int_0^{k_F}
\fr{|t_N|^2}{\eps(k)-\eps_N} \fr{L}{\pi} dk
= \fr{- \Gamma}{2\pi} \ln\left( \fr{4E_F}{\eps_N} \right) .
\ee
Here and everywhere below, $\eps_N$ (as well as e.g. $\eps_1$)
is the function of $V_g$~(\ref{1}). The upperscript $(0)$ at
$E^{(0)}_{tot}$ shows the number of electrons at the narrow
levels (with $i>0$) in the QD. Generalization of this result for
the case of negative $\eps_N$, $\eps_N \ll -\Gamma$ (the broad
level being below the Fermi energy) is straightforward (note
that the level $\eps_N$ is occupied, {\em not the level $\eps_1$
as one might expect}):
\bq\label{5}
E^{(0)}_{tot}= \eps_N -\fr{\Gamma}{2\pi} \ln\left(
\fr{4E_F}{|\eps_N|} \right) \, .
\ee
Here the first term $\eps_N$ accounts for the energy loss due to
replacement of one electron from lead to the dot. The second
order level shift now includes both lowering of levels with
$\eps(k)<\eps_N$ and raising of those with $\eps(k)>\eps_N$. The
perturbative treatment fails for $|\eps(k)-\eps_N|\alt \Gamma$,
but the corresponding shifts of levels below and above $\eps_N$
evidently compensate each other, which is equivalent to taking
the principal value of the integral in Eq~(\ref{3}).  An
approach related to ours was used recently in
ref.~\cite{Kaminski} for the calculation of CB peaks positions.

Finally, the exact solution (for spinless electrons) for a
single state interacting with a continuum is also known (e.g.
\cite{Bohr}). A precise treatment of this situation, along the
lines of ref.\cite{Landau}, yields:
\bq\label{4}
E^{(0)}_{tot} =
\fr{-\Gamma}{4\pi} \left[
\ln\left(\fr{16E_F^2}{\eps_N^2+\Gamma^2/4}\right)
 +2 \right] 
 + \fr{\eps_N}{\pi} 
\cot^{-1}  \fr{2\eps_N}{\Gamma}  ,
\ee
which coincides with the Eqs.~(\ref{3},\ref{5}) at
$|\eps_N|\gg\Gamma$. 

Let us now consider the branch where the level $1$ in the QD is
occupied. The energy of this electron is $\eps_1$. However,
adding one more electron via the hopping $t_N$ {\em now costs
$\eps_N+U_{CB}$}. The ensuing reduction of the downward shift of
the level $E^{(1)}_{tot}$ is of crucial importance. The analog
of Eq.  (\ref{3}) for $\eps_N+U_{CB}>0$ now reads
\bq\label{6}
E^{(1)}_{tot}= \eps_1 -\fr{\Gamma}{2\pi} \ln\left(
\fr{4E_F}{\eps_N+U_{CB}} \right) \, .
\ee

For small $V_g$ one has $E^{(0)}_{tot}<E^{(1)}_{tot}$ and the
Eqs. (\ref{3},\ref{4},\ref{5}) describe the true ground state of
the system. However, the two functions $E^{(0)}_{tot}(V_g)$ and
$E^{(1)}_{tot}(V_g)$ cross at 
\bq\label{8}
\eps_N=-\fr{U_{CB}}{\exp\{ 2\pi(\eps_N -\eps_1)/\Gamma\} +1} 
\ee
and the ground state jumps onto the branch $E^{(1)}_{tot}$. The
energy of the current-transmitting {\it virtual} state $N$ is
positive again. Thus, the transmission amplitude phase had
returned to what it was before the process of filling of state
$N$ and the subsequent sharp jump into the state where level $1$
is filled. It is the latter jump which provides the sharp drop
by $\pi$ of the transmission  phase, following its increase by
$\pi$ through the broad resonance. Many ($\sim (\Gamma/\Delta) 
ln(U_{CB}/\Gamma)$) consecutive resonances are due to the
transition via one and the same level $N$. 

For electrons with spin, the Breit-Wigner-related
formula~(\ref{4}) does not work.  However, far from the
resonance the perturbation theory may still be used (at least
until the temperature is high enough to be away from the Kondo
effect \cite{Kondo}).  We are not able to discuss in detail the
role of spin in this short note.  Still in this case the many
charging events proceed via one and the same broad resonance,
each accompanied by the increase of phase by $\pi$ which is
compensated by the $-\pi$ jump in the valley.

To {\it illustrate} the relevance of the model eq.~(\ref{Ham})
with single strongly coupled level we  performed numerical
simulations for a model QD of a size $l$ with a simple
polynomial potential (a smooth QD coupled to two leads)
\bq\label{v} 
V= -
4x^2\bigl(1-\fr{x}{l}\bigr)^2 + \Bigl(y+ \fr{x^2}{4l}\Bigr)^2
\Bigl(1+8 \bigl(\fr{x}{l}-\fr{1}{2}\bigr)^2 \Bigr) .  
\ee 
Due to the strong mixing of the $x$ and $y$ coordinates the dot
is expected to be nonintegrable, but, similarly to the
experimental geometry~\cite{Heiblum2}, it is approximately
symmetric.  For simulations we considered the QD on the lattice
and used $l=10$ which was equivalent to $50$ lattice spacings.
The kinetic term is given by the standard nearest neighbor
hopping.  Below we present the results of calculations with the
hopping matrix element $\tau=18$ which corresponds to the dot
with $\sim100$ electrons or $\sim200$ if the spin is included
(similar numbers to those in the experiment).  We have used the
potential $V$ of Eq.(\ref{v}) for $0<x<l$. The lead formed by
the potential $V=3y^2$ was attached at $x<0$ and a hard wall at
$x=l$.

Within the energy interval $1.5<\eps<4.7$ only one mode may
propagate along the lead.  The analysis of solutions of the
Schr\"{o}dinger equation within this interval allowed us to find
the positions and widths of quasi-stationary levels in the dot.
As we expected, the widths fluctuate very strongly from level to
level (by many orders of magnitude). In particular the widths of
two levels $\# 102$ and $\# 108$ exceed sufficiently the level
spacing $\Gamma/\Delta\approx 6$ (the number of states doubled
due to spin). The widths of other levels vary from
$\Gamma/\Delta\sim 1$ to $\Gamma/\Delta\sim 10^{-5} - 10^{-6}$. 

The origin of the hierarchy of widths becomes clear from fig.~1,
where we have plotted $|\psi|^2$ in the QD for (real) $\eps$ at
the top of corresponding resonances. The quantized version of
different variants of classical motion may be found on this
figure. The most narrow level $\# 103$ corresponds to a short
stable transverse periodic orbit.  Other broader levels, such as
$\# 96,106$, may be considered as the projections of the
invariant tori corresponding to quasi--periodic classical
motion. This classical trajectory reaches the line $V(x,y)=\eps$
only at few points.  The candidates for chaotic classical motion
(e.g. $\# 110$) also correspond to relatively broad
resonances~\cite{note}. Even in this case only a part of the QD
is covered by the trajectory.  For the most coupled levels $\#
102$ and $\# 108$ the area covered by the trajectory touches the
lead by its corner.

\vspace{-.5cm}
\begin{figure}[t]
\epsfxsize=8.6cm
\epsffile{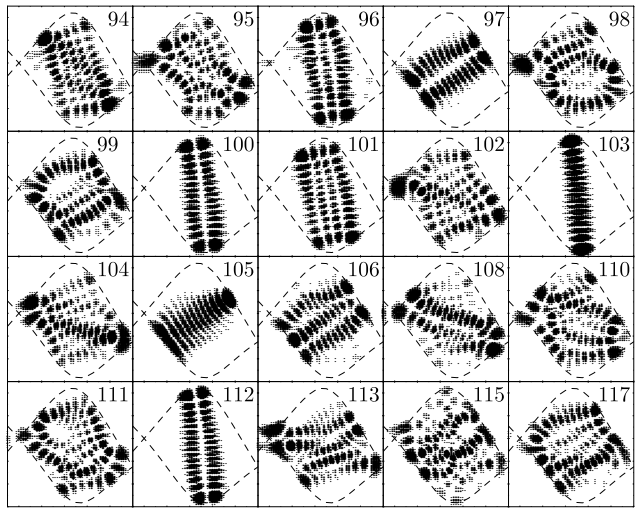}
\vglue 0.2cm
\medskip
\caption{The density 
of electrons in the dot at the resonances coupled to the 
single-channel wire(attached from the left). 
The line $V=0$ is shown dashed. 
Numbers correspond to the number of the level in the QD.
About $95\%$ of the norm of the wave function in the dot is shown.
The ``twin copies'' (such as 100 and 112) of levels 94, 96,
101, 103 are not shown.}
\end{figure}

Moreover, two well coupled trajectories contribute to the level
$\# 102$. This is seen from the fig.~2 where we show also the
$|\psi|^2$ at the left and right wings of this resonance.  One
contribution corresponds to the strongly coupled quasi-periodic
trajectory (left), having the ``turning point'' $V(x,y)=\eps$
just at the left contact.  The other contribution comes from the
true periodic trajectory (right).  Two quantum states in the dot
become mixed via interaction with the wire and form one broad
($\# 102$) and one almost decoupled ($\# 104$)
resonance\cite{Zelevinsky}.

We have repeated the calculations several times for slightly
different $V$ and in a broad range of variation of the hopping.
Typically we saw the resonances of very different width and the
origin of the most broad peaks was explained by simple classical
arguments.


\vspace{-.5cm}
\begin{figure}[t]
\epsfxsize=8.6cm
\epsffile{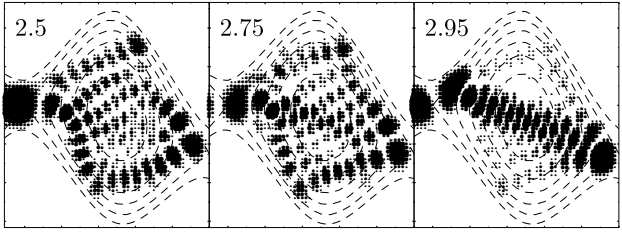}
\vglue 0.2cm
\medskip
\caption{Decomposition of the level $102$ into parts corresponding
to simple classical
trajectories. Numbers are the energies for which the figures
were done.}
\end{figure}

Taking into account the different sensitivities of longitudinal
and transverse modes to the plunger~\cite{Hackenbroich,Gefen}
may allow to keep our broad level $\eps_N$ even longer within
the relevant strip of energy.  This may provide an explanation
of even longer sequences of resonances accompanied by the $-\pi$
jumps.  In a more refined approach, adding new electrons into
the QD should cause a change of the self-consistent potential
$V(x,y)$. The total energy of the dot and the wire will be
lowered in the presence of strongly coupled levels.  This may
cause the potential of the QD to automatically adjust to allow
such levels, which will support our explanation of the
experiment of Ref.~\cite{Heiblum2}.

Our mechanism of charging of the QD requires the existence of
the broad level with $\Gamma\gg\Delta$. The simple way to
justify the relevance of our theory for the explanation of the
experiment of ref.~\cite{Heiblum2} will be to close the dot
sufficiently in order to have $\Gamma\ll\Delta$ for all levels.
In this case the phase still increases by $\pi$ at any
resonance, but the correlation between peaks will disappear.
(More precisely the pairs of peaks corresponding to adding of
electrons with opposite spins onto one and the same level still
are correlated, but correlation between pairs should disappear.)
Moreover, within our mechanism a series of
$\sim(\Gamma/\Delta)\ln (U_{CB}/\Gamma)$ strong charging peaks
in the conductance should have the same height. This ``coupling
dependent'' correlation of the peak heights seems also easy to
measure.

To conclude, we have considered the model, for which upon
increasing $V_g$, it is energetically favorable to first
populate in the dot the level strongly coupled to the leads.  At
a somewhat larger $V_g$ a sharp jump occurs to a state where the
"next in line" narrow level $1$ becomes populated. This jump
accounts for the sharp decrease by $\sim \pi$ of the
transmission phase. The similar  strengths of resonances seen in
the experiment~\cite{Heiblum2} and their large width are also
clear within our mechanism.  The current transmission through
such a QD resembles the behavior of rare earth elements, whose
chemical properties are determined not by the electrons with
highest energy, but by the "strongly coupled" valence electrons.
The overlapping of single-particle resonances may take place
also in the Kondo experiments in QD-s \cite{Kondo1-3}, where in
order to increase the Kondo temperature the dot is usually
sufficiently opened. Hopefully the unusual effects observed in
some of these experiments  may be also explained within our
approach.

Valuable discussions with E.~Buks, M.~Heiblum, Y.~Gefen,
I.~Lerner, Y.~Levinson, M.~Schechter, V.~V.~Sokolov,
D.~Sprinzak, H.~A.~Weidenmuller and A.~Yacoby are acknowledged.
The work of PGS was supported by RFBR, grant 98-02-17905 . Work
at WIS was supported by the Albert Einstein Minerva Center for
Theoretical Physics and by grants from the German-Israeli
Foundation (GIF), the Israel Science Foundation, Jerusalem and
the Tekla and Simon Bond Fund for Submicron Electronics Research
at the Weizmann Institute.


\begin{thebibliography}{99}
\bibitem{note} In order to find the convincing evidence of
``chaos'' one should consider smaller $\hbar$ (more electrons in
the QD).
\end{thebibliography}
\end{document}